\documentclass[a4paper,showpacs,prb,twocolumn]{revtex4}%
\usepackage{amsfonts}
\usepackage{graphicx}
\usepackage{color}
\usepackage{amsmath}
\usepackage{amssymb}
\usepackage{latexsym}
\usepackage{psfrag}%
\providecommand{\U}[1]{\protect\rule{.1in}{.1in}}
\providecommand{\U}[1]{\protect\rule{.1in}{.1in}}
\begin{document}
\title{Quantum wires in magnetic field: A comparative study of the Hartree-Fock and
the spin density functional approaches}
\author{S. Ihnatsenka}
\affiliation{Solid State Electronics, Department of Science and Technology (ITN),
Link\"{o}ping University, 60174 Norrk\"{o}ping, Sweden}
\author{I. V. Zozoulenko}
\affiliation{Solid State Electronics, Department of Science and Technology (ITN),
Link\"{o}ping University, 60174 Norrk\"{o}ping, Sweden}
\date{\today}

\begin{abstract}
We present a detailed comparison of the self-consistent calculations based on the
Hartree-Fock and the spin density functional theory for a spit-gate quantum wire in the
IQH regime. We demonstrate that both approaches provide qualitatively (and in most cases
quantitatively) similar results for the spin-resolved electron density, spin
polarization, spatial spin separation at the edges and the effective $g$ factor. The both
approach give the same values of the magnetic fields corresponding to the successive
subband depopulation and qualitatively similar evolution of the magnetosubbands.
Quantitatively, however, the HF and the DFT subbands are different (even though the
corresponding total electron densities are practically the same). In contrast to the HF
approach, the DFT calculations predict much larger spatial spin separation near the wire
edge for the low magnetic fields (when the compressible strips for spinless electrons are
not formed yet). In the opposite limit of the large fields, the Hatree-Fock and the DFT
approaches give very similar values for the spatial spin separation.

\end{abstract}

\pacs{73.21.Hb, 73.43.Cd, 73.23.Ad}
\maketitle

%\section{Introduction}

\textit{Introduction.} A detailed knowledge of energetics, spin splitting,
magnetosubband and edge state structure in quantum wires is necessary for
understanding and interpretation of a variety of magnetotransport phenomena in
the integer quantum Hall (IQH) regime. A powerful tool to study
electron-electron interaction and spin effects in quantum wires is the
mean-field approaches such as the Hartree-Fock (HF) and the spin density
functional theory (DFT)\cite{Giuliani_Vignale}. A number of studies based on
these approaches addressing various aspects of interacting electrons in the
IQH regime in quantum wires have been reported
recently\cite{Kinaret,Dempsey,Tokura,Manolescu,DFT_fractional,Stoof,Takis,Brataas,Takis2,Balev,Struck,Ihnatsenka1,Ihnatsenka2,Ihnatsenka34}%
. However, in some cases different studies arrive to conflicting results and
findings reported in some studies are not recovered in others. It is not clear
whether the reason for such discrepancies is due to utilization of different
approaches treating the exchange and correlation effects in different ways
(i.e. HF vs spin DFT), or this difference is related to various approximations
of different models (such as e.g. neglecting a global electrostatics,
simplified models for screening, non self-consistent calculations, fixed
filling factors, etc.).

The aim of this Brief report is to resolve this issue by presenting a detailed
comparison of the self-consistent calculations based on the HF method and the
spin DFT approximation for a spit-gate quantum wire in the IQH regime. This
includes a comparison of the magnetosubband structure, electron densities,
spin polarization and spatial spin separation as well as calculation of the
effective $g$ factor. We stress that in our calculations we do not use any
simplified assumptions concerning screening such that a global electrostatics
of the system at hand is treated in an exact way. Note that comparative
studies of different approaches are common in treatment of electronic
properties of quantum dots as they provide an important insight into the
validity of applied methods and used approximations\cite{QDOverview}. At the
same time, we are not aware of corresponding studies for the quantum wires in
the IQH regime. Another motivation for the present study is that the DFT based
approaches are often used for transport calculations. It has been argued
recently that the standard DFT approaches might not be always suitable for
this purpose (notably in the weak coupling regime), because of the inherent
problems of the derivative discontinuity problem and related uncompensated
self-interaction errors in the DFT\cite{SIE}. At the same time, the HF
approach does not suffer from the above problems and thus the utilization of
the HF method in the transport calculations overcomes the limitations of the
standard DFT. Thus, the detailed comparison of these two methods can provide a
justification for the utilization of the HF instead of the DFT.

%\section{Basics}

\textit{Basics} We consider an infinitely long split-gate GaAs/AlGaAs quantum
wire in a perpendicular magnetic field $B$ where electrons are situated at the
distance $b$ below the surface. The HF equation for a single-particle wave
function of spin $\sigma,$ $\Phi_{\beta}^{\sigma}(\mathbf{r}),$
reads\cite{Giuliani_Vignale}%
\begin{align}
&  \left[  H_{0}(\mathbf{r})+V_{conf}(y)+V_{H}(y)+V_{Z}\right]  \Phi_{\beta
}^{\sigma}(\mathbf{r})\nonumber\\
&  +\int V_{Fock}(\mathbf{r},\mathbf{r}^{\prime})\Phi_{\beta}(\mathbf{r}%
^{\prime})\,d\mathbf{r}^{\prime}=E_{\beta}\Phi_{\beta}^{\sigma}(\mathbf{r}%
),\label{HF1}%
\end{align}
where $\mathbf{r}=(x,y)$, $H_{0}(\mathbf{r})=-\frac{\hbar^{2}}{2m^{\ast}%
}\left(  \frac{\partial}{\partial x}-\frac{eiBy}{\hbar}\right)  ^{2}%
+\frac{\partial^{2}}{\partial y^{2}}$ is the kinetic energy in the Landau
gauge, with $m^{\ast}=0.067m_{e}$ being the GaAs effective mass; $\sigma
=\pm\frac{1}{2}$ describes spin-up and spin-down states, $\uparrow$ ,
$\downarrow$. In the split-gate geometry the bare confining potential
$V_{conf}(y)$ due to the gates, donor layers and the Schottky barrier is well
approximated by the parabolic confinement, $V_{conf}(y)=V_{0}+\frac{m^{\ast}%
}{2}\left(  \omega_{0}y\right)  ^{2}$, where $V_{0}$ defines the bottom of the
potential (we set the Fermi energy $E_{F}=0)\cite{Ihnatsenka1}.$ The Zeeman
energy is $V_{Z}=g\mu_{b}B\sigma$ where $\mu_{B}=e\hbar/2m_{e}$ is the Bohr
magneton and the bulk $g$ factor of GaAs is $g=-0.44.$ The Hartree potential
due to the electron density $n(y)=\sum_{\sigma}n^{\sigma}(y)$ (including the
mirror charges) is$\cite{Ihnatsenka1}$ $V_{H}(y)=-\frac{e^{2}}{4\pi
\varepsilon_{0}\varepsilon_{r}}\int dy^{\prime}n(y^{\prime})\ln\frac{\left(
y-y^{\prime}\right)  ^{2}}{\left(  y-y^{\prime}\right)  ^{2}+4b^{2}}.$ The
non-local Fock operator is $V_{Fock}(\mathbf{r},\mathbf{r}^{\prime}%
)=-\frac{e^{2}}{4\pi\varepsilon\varepsilon_{0}|\mathbf{r}-\mathbf{r}^{\prime
}|}\sum_{\beta}f_{E_{\beta}}^{FD}\Phi_{\beta}^{\sigma}(\mathbf{r})\Phi_{\beta
}^{\sigma\ast}(\mathbf{r}^{\prime})$, where the summation is performed over
all states $\beta,$ and $f_{E_{\beta}}^{FD}$ is the Fermi-Dirac distribution function.

We assume the Bloch form of the wave function,
\begin{equation}
\Phi_{n,k}^{\sigma}(x,y)=e^{ikx}\varphi_{n,k}^{\sigma}(y)\label{Bloch}%
\end{equation}
where $k$ is the wave vector, $\varphi_{n,k}(y)^{\sigma}$ describes the $n$-th
transverse subband for the spin $\sigma$. Substituting the Bloch function
(\ref{Bloch}) into the HF equation (\ref{HF1}) and integrating over the
longitudinal coordinate $x$ we arrive to the set of eigenequations for
$\varphi_{n,k}^{\sigma}(y),$ \cite{Brataas}%
\begin{align}
&  \bigg[-\frac{\hbar^{2}}{2m^{\ast}}\frac{d^{2}}{dy^{2}}+\frac{m^{\ast}%
\omega_{c}^{2}}{2}\left(  y+\frac{\hbar k}{eB}\right)  ^{2}+V_{conf}%
(y)+V_{H}(y)\nonumber\\
&  +V_{Z}\bigg]\varphi_{n,k}^{\sigma}(y)+\int V_{Fock}^{k}(y,y^{\prime
})\varphi_{n,k}^{\sigma}(y^{\prime})\,dy^{\prime}=E_{n,k}^{\sigma}%
\varphi_{n,k}^{\sigma}(y),\label{HF2}%
\end{align}
where $\omega_{c}$ is the cyclotron frequency and
\begin{align}
V_{Fock}^{k}(y,y^{\prime}) &  =-\frac{e^{2}}{2\pi\varepsilon_{0}%
\varepsilon_{r}}\sum_{n^{\prime}k^{\prime}}f_{E_{n,k}^{\sigma}}^{FD}%
\varphi_{n^{\prime}k^{\prime}}^{\sigma}(y)\varphi_{n^{\prime}k^{\prime}%
}^{\sigma\ast}(y^{\prime})\nonumber\\
&  \times K_{0}\left(  |k-k^{\prime}||y-y^{\prime}|\right)  ,
\end{align}
with $K_{0}$ being the modified Bessel function. Discrediting Eq. (\ref{HF2})
we reduce the system of the integro-differential equations to the system of
linear equations, which we solve numerically by standard methods in an
iterative way until the self-consistent solution is achieved. Knowledge of the
wave vectors $k_{n}^{\sigma}$ for different states $\{n,\sigma\}$ allows us to
recover the subband structure\cite{Ihnatsenka1}, i.e. to calculate an overage
position $y_{n}^{\sigma}$ of the wave functions for different modes $n$,
$y_{n}^{\sigma}=\hbar k_{n}^{\sigma}/eB.$

Within the framework of the spin density functional theory, the Kohn-Sham
equations for the single particle wave function $\Phi_{\beta}^{\sigma
}(\mathbf{r})$ read
\begin{align}
&  \left[  H_{0}(\mathbf{r})+V^{\sigma}(y)\right]  \Phi_{\beta}^{\sigma
}(\mathbf{r})=E_{\beta}\Phi_{\beta}^{\sigma}(\mathbf{r}),\label{KS}\\
&  V^{\sigma}(y)=V_{conf}(y)+V_{H}(y)+V_{Z}+V_{xc}^{\sigma,\zeta}(y),
\label{V}%
\end{align}
where the first three terms in the effective confinement potential $V^{\sigma
}(y)$ are the same as in the HF equation (\ref{HF1}), and the last term
corresponds to the exchange and correlation potential in the local spin
density approximation. It is given by the functional derivative $V_{xc}%
^{\sigma,\zeta}=\frac{\delta}{\delta n^{\sigma}}\left\{  n\epsilon_{xc}%
^{\zeta}\left(  n\right)  \right\}  ,$ where $\epsilon_{xc}^{\zeta}\left( n\right)  $ is
the exchange and correlation energy functional and
$\zeta(y)=\frac{n^{\uparrow}-n^{\downarrow}}{n^{\uparrow}+n^{\downarrow}}$ is the local
spin polarization. All the results presented below correspond to the parameterization of
$\epsilon_{xc}^{\zeta}\left(  n\right)  $ given by Tanatar and Ceperley\cite{TC}.
Assuming the Bloch form of the wave functions (\ref{Bloch}), the equations (\ref{KS}) are
solved self-consistently using the Green's function technique as described in detail in
Ref. \onlinecite{Ihnatsenka1} (see also Refs.
\onlinecite{Ihnatsenka2},\onlinecite{Ihnatsenka34}).

Note that we find the self-consistent solutions for the DFT and the HF approaches using
completely unrelated numerical methods. As a validity check we control that these
different methods give identical results when we set $V_{Fock}^{k}$ and
$V_{xc}^{\sigma,\zeta}$ in respectively Eqs. (\ref{HF2}) and (\ref{V}) to zero and thus
reduce both approaches to the standard spinless Hartree approximation (the latter is
shown to reproduce well\cite{Ando,Ihnatsenka2} the Chklovskii \textit{et
al.}\cite{Chklovskii} electrostatic treatment).

%*********************************************************
%*********************************************************
\begin{figure}[ptb]
\includegraphics[scale=0.9]{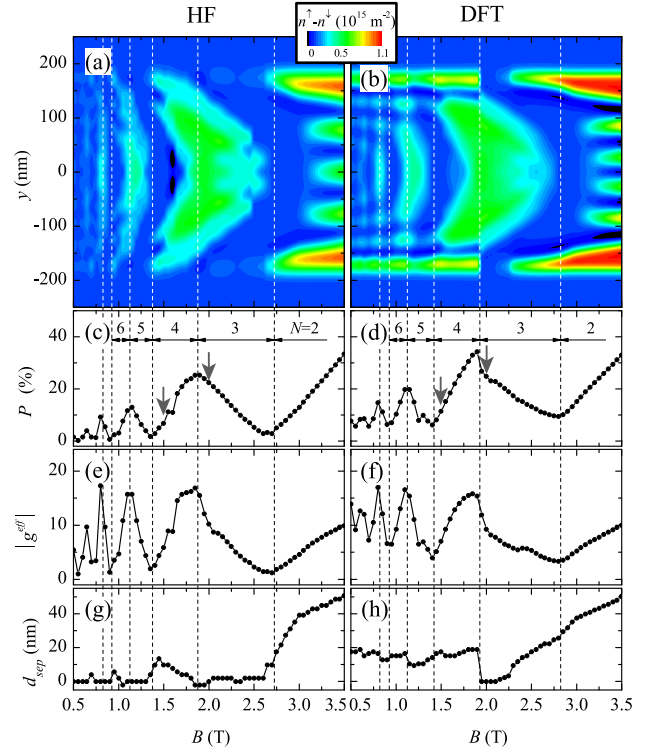}
%[tph]
\caption{(Color online). (a),(b) Spatially resolved spin polarization of the electron
density $n^{\uparrow}-n^{\downarrow}$. (c),(d) The number of subbands and the total spin
polarization $P=\frac{n_{1D}^{\uparrow}-n_{1D}^{\downarrow
}}{n_{1D}^{\uparrow}+n_{1D}^{\downarrow}}$; arrows indicate the magnetic fields
corresponding to the magnetosubband structure shown in Fig. \ref{fig:band_diagram};
(e),(f) The effective $g$ factor. ( g),(h) The spatial spin separation at the wire edge
$d_{sep}$ (a definition of $d_{sep}$ is outlined in Fig. \ref{fig:band_diagram} (a) ).
The left and right panels correspond respectively to the HF and the spin DFT
approximations. The bare confining potential $V_{conf}(y)$ is well approximated by the
parabolic confinement with $V_{0}=-85$meV, $\hbar\omega_{c}=2$meV, the distance to the
surface
$b=60$nm, temperature $T=1$K.}%
\label{fig:g-factor}%
\end{figure}
%*********************************************************
%*********************************************************

%\section{Results and discussion}

\begin{figure}[ptb]
\includegraphics[scale=0.9]{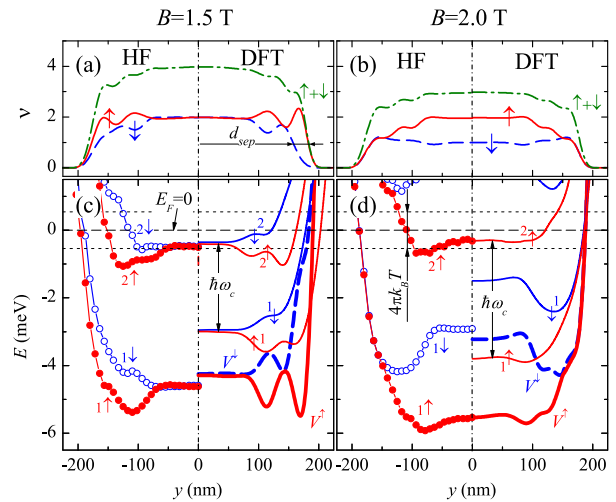}
%[tph]
\caption{(Color online). (a),(b) The local filling factors $\nu^{\uparrow}%
,\nu^{\downarrow}$ and $\nu=\nu^{\uparrow}+\nu^{\downarrow}$ ($\nu=nh/eB$)
calculated within the HF and the spin DFT approaches for two representative
magnetic fields (indicated by arrows in Fig. \ref{fig:g-factor} (c),(d).
(c),(d) The magnetosubband structure for spin-up and spin-down electrons
calculated within the HF and the spin DFT approaches. The fat solid lines
indicate the DFT effective confinement potential $V^{\sigma}$, Eq. (\ref{V})
(note that because of the nonlocal character of the HF equations, it is not
possible to define the effective confinement potential in the HF approach.)}%
\label{fig:band_diagram}%
\end{figure}

\textit{Results and discussion}. Figure \ref{fig:g-factor} show a spatially
resolved difference in the electron density $n^{\uparrow}-n^{\downarrow}$ and
the total spin polarization $P=\frac{n_{1D}^{\uparrow}-n_{1D}^{\downarrow}%
}{n_{1D}^{\uparrow}+n_{1D}^{\downarrow}}$ calculated using the HF and the spin DFT
approaches for a representative quantum wire ($n_{1D}^{\sigma}=\int n^{\sigma}(y)dy$). A
detailed analysis of the spin polarization in a split-gate quantum wire based on the
spin-DFT approach is given in Refs. \onlinecite{Ihnatsenka1,Ihnatsenka2}. For the sake of
comparison with the HF approximation we summarize below the main finding. The spin
polarization is maximal for magnetic fields close to the depopulation of the even
subbands, see Fig. \ref{fig:g-factor} (d). In this case the highest occupied (odd)
subband forms a compressible strip in the middle of the wire, and, therefore, the
electron density is mostly spin-polarized in the centre of the wire, see Fig.
\ref{fig:band_diagram} (d). (We define the width of the compressible
strip within the window $|E-E_{F}|\leqslant2\pi kT$%
,\cite{Ihnatsenka1,Ihnatsenka2,Ihnatsenka34,Ando} which corresponds to the
energy interval where the subbands are partially filled, $0<f^{FD}<1$). Minima
of the spin polarization correspond, instead, to depopulation of the odd
subbands. At polarization minima the spin-up and spin-down subbands are fully
(and practically equally) occupied as their bottoms in the center of the wire
are situated below (or just on the border) of the energy window $|E-E_{F}%
|\leqslant2\pi kT$, see Fig. \ref{fig:band_diagram} (c). Because of this, the
spin polarization in the centre of the wire is absent and it increases toward
the edges of the wire because the spin-up and spin-down subbands intersect
$E_{F}$ at different distances from the wire center. For the case of spinless
electrons the compressible strips are formed near the wire boundaries for
sufficiently high magnetic fields\cite{Chklovskii,Ihnatsenka2,Ando}. The
exchange interaction, however, completely or partially suppress the
compressible strips leading to a spatial spin separation between the spin-up
and spin-down states\cite{Ihnatsenka2}. This spatial separation causes a
strong spin polarization near the boundaries which is clearly seen in Fig.
\ref{fig:g-factor} (b) for magnetic fields $B\gtrsim2.75$ T. This spin
separation grows as the magnetic field increases because the width of the
corresponding compressible strips for the spinless electons increases, see
Figs. \ref{fig:g-factor} (b), (h). Note that the spin-DFT approach also
predicts an almost constant (independent of $B$) spatial spin polarization
even for lower fields $B\lesssim2.0$ T, when the compressible strips are not
formed yet.
%*********************************************************
%*********************************************************
%*********************************************************
%*********************************************************
%*********************************************************
%*********************************************************
\begin{figure}[b]
\includegraphics[scale=0.8]{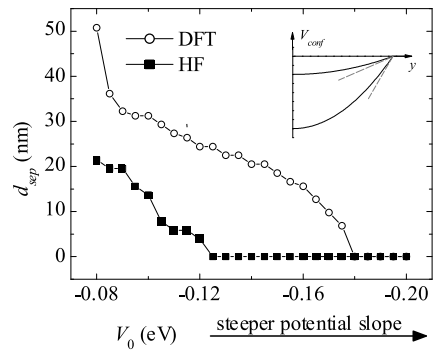}
%[tph]
\caption{The spatial spin separation near the wire edge as a function of the confinement
steepness calculated within the spin DFT and HF approaches in the regime of low fields
($B=0.5$T). The inset illustrates how the confinement steepness changes as the bottom of
the parabolic confinement potential, $V_{0}$, varies ($\hbar\omega_0$ is adjusted to keep
the wire width
constant, $w=500$nm.) }%
\label{fig:separation}%
\end{figure}
%*********************************************************
%*********************************************************

Let us now compare the results of the spin-DFT calculations with those based on the HF
approach. It is remarkable that the both approaches give practically the same values of
the magnetic fields corresponding to the successive subband depopulation, c.f. Figs.
\ref{fig:g-factor} (c) and (d). Moreover, both approaches give practically the same
\textit{total} electron density distribution $n(y),$ see Fig. \ref{fig:band_diagram} (a)
and (b). However, the \textit{spin-resolved} densities are not always the same. In
contrast to the spin-DFT approach, the HF calculation does not always exhibit a spatial
spin polarization near the edges for the low fields (when the compressible strips for
spinless electrons are not formed yet). This is the case for a quantum wire of Fig.
\ref{fig:g-factor} for $B\lesssim2.75$ T. The spin separation near the wire edges
$d_{sep}$ is caused by the exchange interaction and it is known to depend on the
steepness of the confinement potential\cite{Dempsey,Muller}: as the external confining
potential becomes smoother, the spin separation growths. Figure \ref{fig:separation}
shows that while the spatial spin separation $d_{sep}^{DFT}$ and $d_{sep}^{HF}$ exhibit
qualitatively same behavior as a function of the potential steepness, the DFT approaches
predicts much larger spatial spin separation as compared to the HF method. Besides, the
critical value of the potential steepness at which different spins become spatially
separated is obviously lower in the HF approach. We stress that the difference between
$d_{sep}^{DFT}$ and $d_{sep}^{HF}$ discussed above corresponds to the regime of the low
fields, when the compressible strips for spinless electrons are not formed yet. For
larger fields (corresponding to the formation of compressible strips for the spinless
electrons), the Hatree-Fock and the DFT approaches give very similar values for the spin
separation, c.f. Figs. \ref{fig:g-factor} (g) and (h). In this case $d_{sep}$ is
approximately equal to the width of the compressible strips for spinless electrons (see
Ref. \onlinecite{Ihnatsenka2} for a detailed discussion of the suppression of the
compressible strips by the exchange interaction leading to the spatial spin polarization
at the edges). Note that our study can not distinguish which approach gives a correct
result for $d_{sep}$ for the low field. This question can be resolved by a comparison to
the exact results obtained by e.g. quantum Monte Carlo methods. We speculate at this
point that it is the DFT approach that overestimates the spatial spin separation at lower
fields. This conclusion is based on transport measurements on lateral quantum dots
indicating that the spin polarized injection and detection by means of the spatial
separation of spins can be achieved only in the edge state regime for sufficiently high
magnetic field\cite{Andy}.

Let us turn our attention to the subband structure. Figure
\ref{fig:band_diagram} shows the subband structure for two representative
magnetic fields corresponding to the minimum and maximum of the spin
polarization in a quantum wire. Qualitatively, the HF and the DFT subbands
exhibit very similar features and evolve in a similar way as the magnetic
field is changed. This includes the subband depopulation, the formation of the
compressible strip in the middle of the wire and the subband separation at the
edges. Quantitatively, however, the HF and the DFT subbands are different
(even though the corresponding densities are practically the same, see Fig.
\ref{fig:band_diagram} (b), (d)). The most pronounced difference is that the
consecutive subband separation for different spins in the DFT approach is
equal to $\hbar\omega_{c}$, whereas the HF subband separation exceeds this
value. We attribute this difference to the nonlocal character of the exchange
interaction in the HF approximation. Note that the HF subband separation tends
to $\hbar\omega_{c}$ as the density increases because the exchange interaction
becomes less pronounced for higher densities in comparison to the kinetic energy.

Because the DFT and HF approaches give rather similar evolution of the
magnetosubband structure, the corresponding behavior of the total
spin-polarization $P$ and the effective $g$ factor, $g^{eff}$, is also
similar, see Figs. \ref{fig:g-factor} (c), (d), and (e), (f). (We define the
effective $g$ factor according to $g^{eff}=\left\langle (E_{n,k}^{\uparrow
}-E_{n,k}^{\downarrow})/g\mu_{B}B\right\rangle $ where the averaging is
performed over the all $k$ vectors and the occupied subbands $n$). The DFT
approach gives a slightly higher value of $P$ at the lower fields because of
the enhanced spin polarization near the edges as discussed above. The both
approaches give quantitatively similar dependencies of $g^{eff}$ as a function
of magnetic field. Because $g^{eff}$ is directly related to the subband spin
splitting, the dependence of $g^{eff}=g^{eff}(B)$ closely follows that of
$P=P(B)$, showing a well-known oscillatory character with a periodicity of
$1/B$ related to the subband depopulation\cite{AFS}. A maximum value of
$g^{eff}\approx15$ is reached close to magnetic fields corresponding to
depopulation of the even subbands, i.e. when the subband splitting in the wire
center is maximal.

%\section{Conclusion}
\textit{Conclusion.} We demonstrate that the spin DFT and the HF approaches
provide qualitatively (and in most cases quantitatively) same description of a
split-gate quantum wire in the IQH regime. This includes the electron density,
spin polarization and the effective $g$ factor. The both approach give the
same values of the magnetic fields corresponding to the successive subband
depopulation and qualitatively similar evolution of the magnetosubbands.
Quantitatively, however, the HF and the DFT subbands are different (even
though the corresponding total electron densities are practically the same).
In contrast to the HF approach, the DFT calculations predict much larger
spatial spin separation near the wire edge for the low fields (when the
compressible strips for spinless electrons are not formed yet).

\end{document}